\documentclass{PoS}
\usepackage{amsmath}
\usepackage{amssymb}
\usepackage{xspace}
\usepackage{nicefrac}

\newcommand{\Comix}{\textsc{Comix}\xspace}
\newcommand{\Amegic}{\textsc{Amegic}\xspace}
\newcommand{\Collier}{\textsc{Collier}\xspace}
\newcommand{\CutTools}{\textsc{CutTools}\xspace}
\newcommand{\OneLoop}{\textsc{OneLOop}\xspace}
\newcommand{\Munich}{\textsc{Munich}\xspace}
\newcommand{\Sherpa}{\textsc{Sherpa}\xspace}
\newcommand{\OpenLoops}{\textsc{OpenLoops}\xspace}
\newcommand{\SherpaOpenLoops}{\Sherpa{}+\OpenLoops\xspace}
\newcommand{\MunichOpenLoops}{\Munich{}+\OpenLoops\xspace}

\newcommand{\MCatNLO}{M\protect\scalebox{0.8}{C}@N\protect\scalebox{0.8}{LO}\xspace}

\newcommand{\MEPSatNLO}{M\protect\scalebox{0.8}{E}P\protect\scalebox{0.8}{S}@N\protect\scalebox{0.8}{LO}\xspace}

\newcommand{\mr}[1]{\mathrm{#1}}

\newcommand{\order}{\mathcal{O}}
\newcommand{\alphaS}{\alpha_s}
\newcommand{\Qcut}{Q_\text{cut}}
\newcommand{\rT}{\mr{T}}
\newcommand{\pT}{p_\rT}

\title{NLO QCD+EW for V+jets}

\ShortTitle{NLO QCD+EW for V+jets}

\author{\speaker{Marek Sch\"onherr}\\
        Physik-Institut, Universit\"at Z\"urich,
	Winterthurerstrasse 190, 
	CH-8057 Z\"urich,
	Switzerland\\
        E-mail: \email{marek.schoenherr@physik.uzh.ch}}

\author{Stefan Kallweit\\
        Institut f\"ur Physik \& \textsc{Prisma} Cluster of Excellence, 
        Johannes Gutenberg Universit\"at, 55099 Mainz, Germany\\
        E-mail: \email{kallweit@uni-mainz.de}}

\author{Jonas M.\ Lindert\\
        Physik-Institut, Universit\"at Z\"urich,
	Winterthurerstrasse 190, 
	CH-8057 Z\"urich,
	Switzerland\\
        E-mail: \email{lindert@physik.uzh.ch}}

\author{Philipp Maierh\"ofer\\
        Physikalisches Institut, 
        Albert-Ludwigs-Universit\"at Freiburg,
        79104 Freiburg, Germany\\
        E-mail: \email{philipp.maierhoefer@physik.uni-freiburg.de}}

\author{Stefano Pozzorini\\
        Physik-Institut, Universit\"at Z\"urich,
	Winterthurerstrasse 190, 
	CH-8057 Z\"urich,
	Switzerland\\
        E-mail: \email{pozzorin@physik.uzh.ch}}

\abstract{In this contribution recent results regarding the NLO
   electroweak corrections for vector boson production in association with
   jets are presented. Besides discussing the phenomenology of the 
   fixed-order results, their incorporation in existing NLO QCD parton
   shower matched and merged calculations, which can directly be used in
   experimental analyses, will be shown.}

\FullConference{Fourth Annual Large Hadron Collider Physics\\
		13-18 June 2016\\
		Lund, Sweden}

\begin{document}

\section{Introduction}
\label{Sec:intro}

The production of electroweak gauge bosons, frequently accompanied by 
one or multiple jets, plays a key role in the physics programme 
of the Large Hadron Collider. When decaying into leptons, the comparatively 
clean final state offers unique opportunities to test the Standard Model 
at the highest precisions. Simultaneously, vector boson production in 
association with multiple jets represents a large and therefore 
important background to many new physics searches. 

As predictions for vector boson plus jets at next-to leading order (NLO) 
in QCD \cite{Arnold:1988dp,Arnold:1989ub,Campbell:2002tg,Berger:2009zg,
  Ellis:2009zw,Berger:2010zx,Bern:2013gka} are widely available, and the 
calculation of $V+1\,\text{jet}$ final states is even known at next-to-next-to 
leading order (NNLO) \cite{Boughezal:2016isb,Boughezal:2016dtm,
  Boughezal:2016yfp,Boughezal:2015dva,Ridder:2016nkl,Ridder:2015dxa},  
electroweak (EW) corrections become increasingly important. They are relevant both for 
precision observables and measurements in the TeV regime where large Sudakov 
logarithms can lead to sizeable reductions of cross sections
\cite{Fadin:1999bq,Kuhn:1999nn,Denner:2000jv,Denner:2001gw,Ciafaloni:2000df,
  Mishra:2013una}. While specific calculations for low multiplicity 
processes were available for some time \cite{Kuhn:2005gv,Kuhn:2007cv,
  Denner:2009gj,Denner:2011vu,Denner:2012ts}, the recent 
automation of the generation of one-loop scattering amplitudes delivered 
the flexibility to calculate NLO EW corrections for processes with multiple 
jets in the final state \cite{Denner:2014ina,Kallweit:2014xda,
  Kallweit:2015dum,Chiesa:2015mya}. Automated implementations of the pure 
virtual Sudakov corrections are also available \cite{Chiesa:2013yma}.

\section{Next-to-leading order electroweak corrections}

\begin{figure}[t]
  \includegraphics[width=0.47\textwidth]{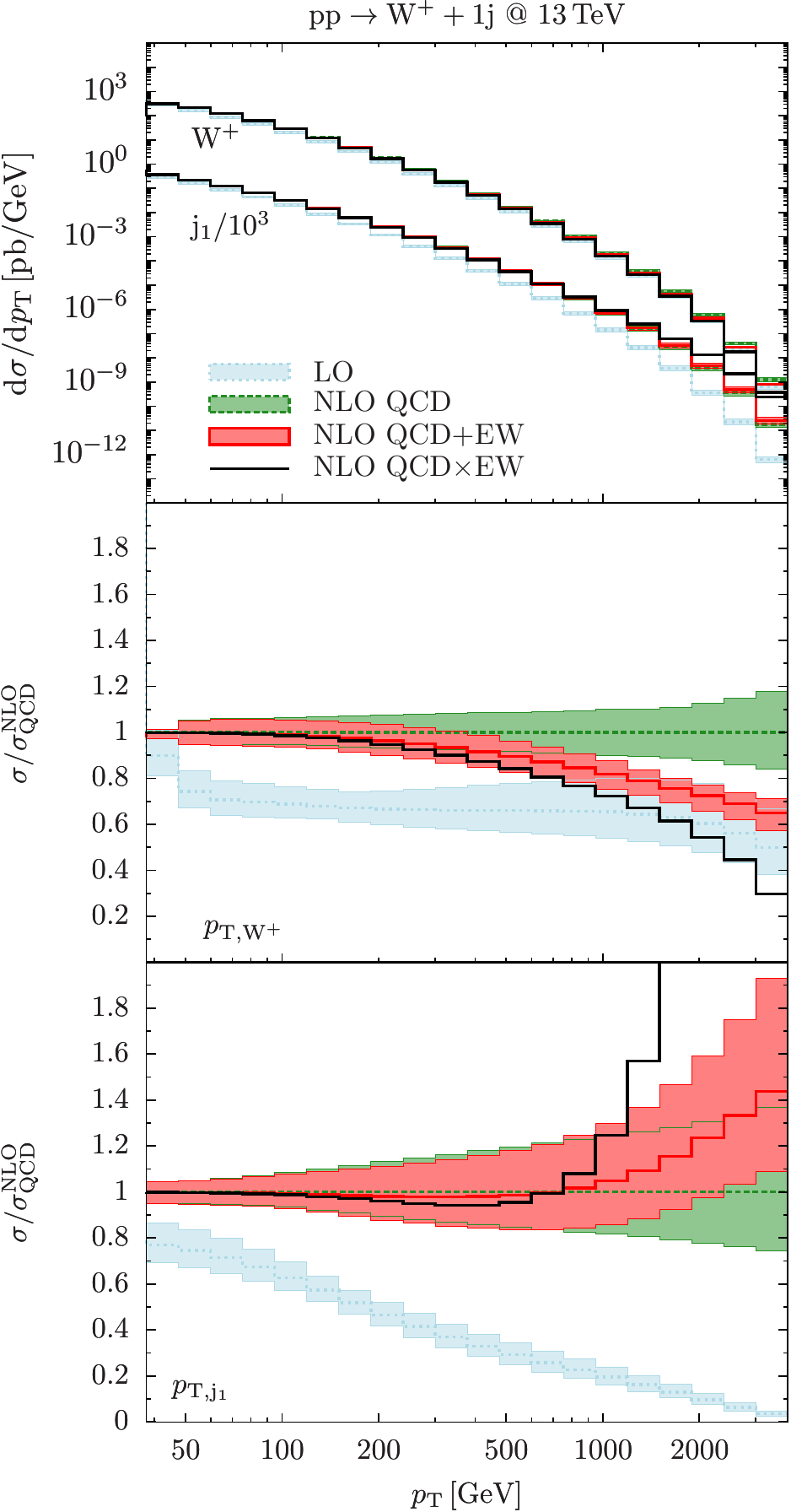}\hfill
  \includegraphics[width=0.47\textwidth]{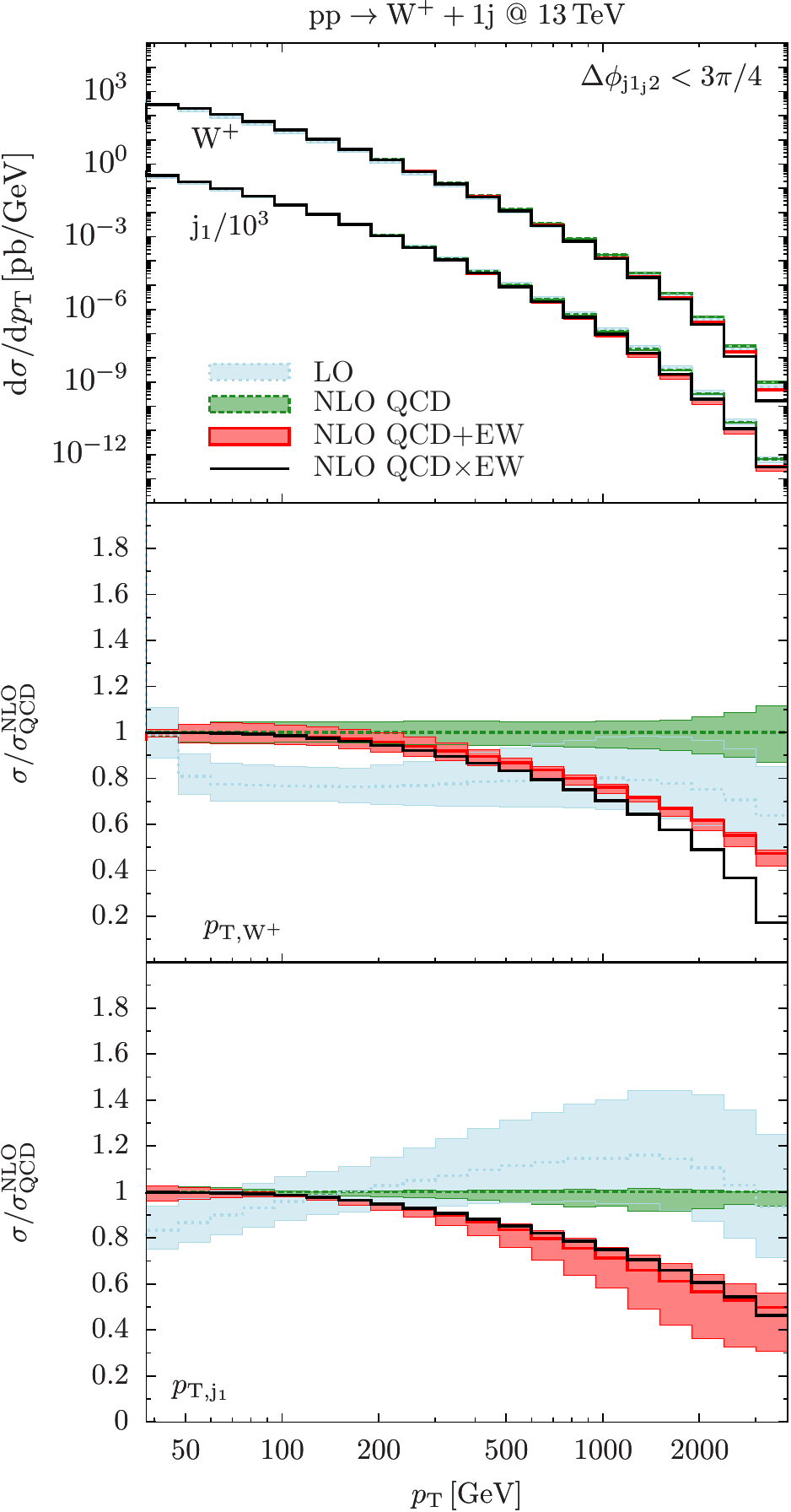}
  \caption{
    \label{Fig:wj}
    NLO QCD and EW corrections to the $W$-boson and leading jet 
    transverse momenta in $pp\to W^+j$ at 13\,TeV at the LHC. 
    While both types of corrections are dominated by real radiation 
    $pp\to W^+jj$ topologies featuring back-to-back dijet topologies 
    and subsequently atypically large $K$-factors in the inclusive 
    case (left), vetoing such configurations (right) recovers the 
    standard behaviour.
  }
\end{figure}

\begin{figure}[t]
  \includegraphics[width=0.47\textwidth]{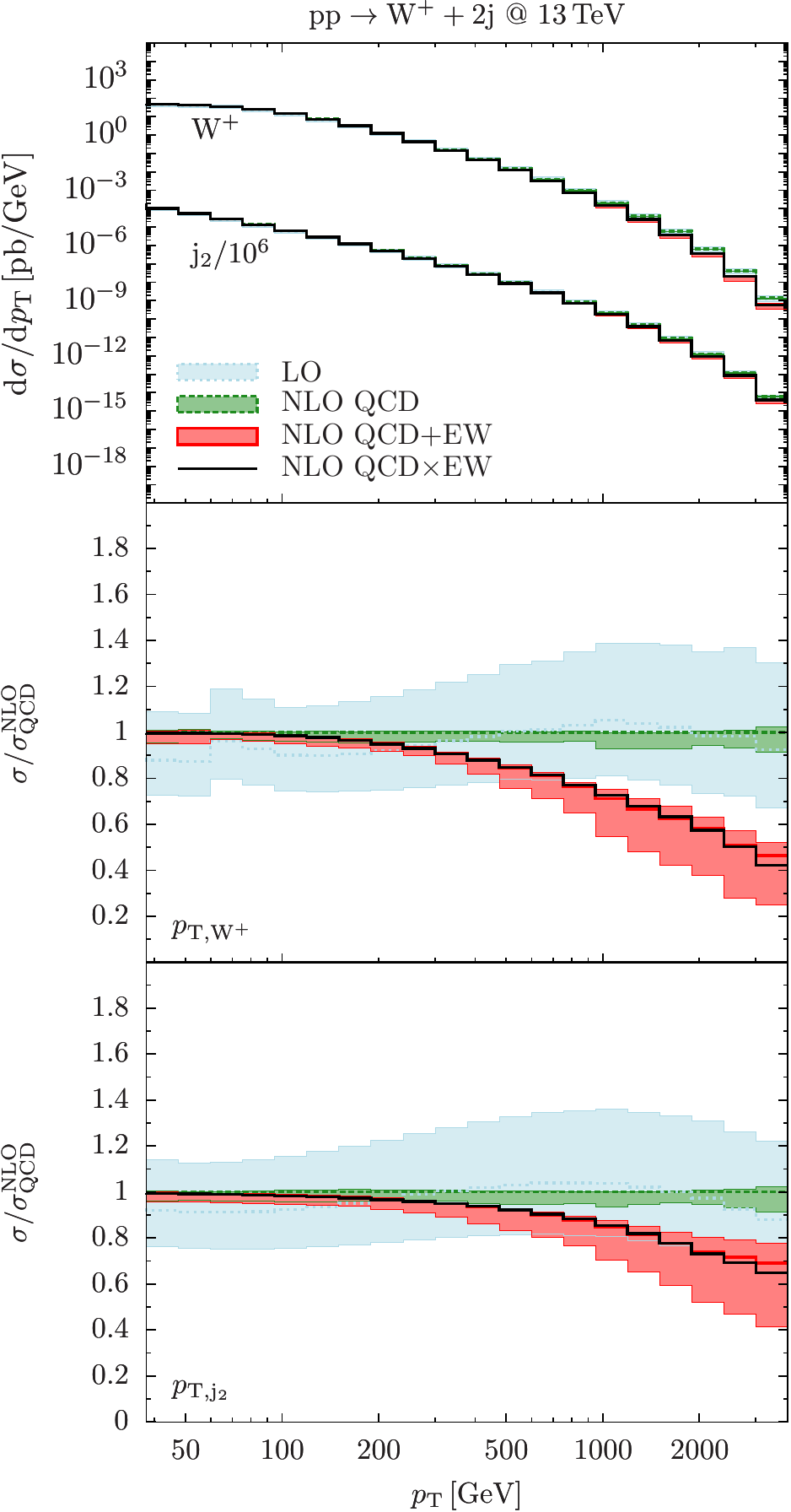}\hfill
  \includegraphics[width=0.47\textwidth]{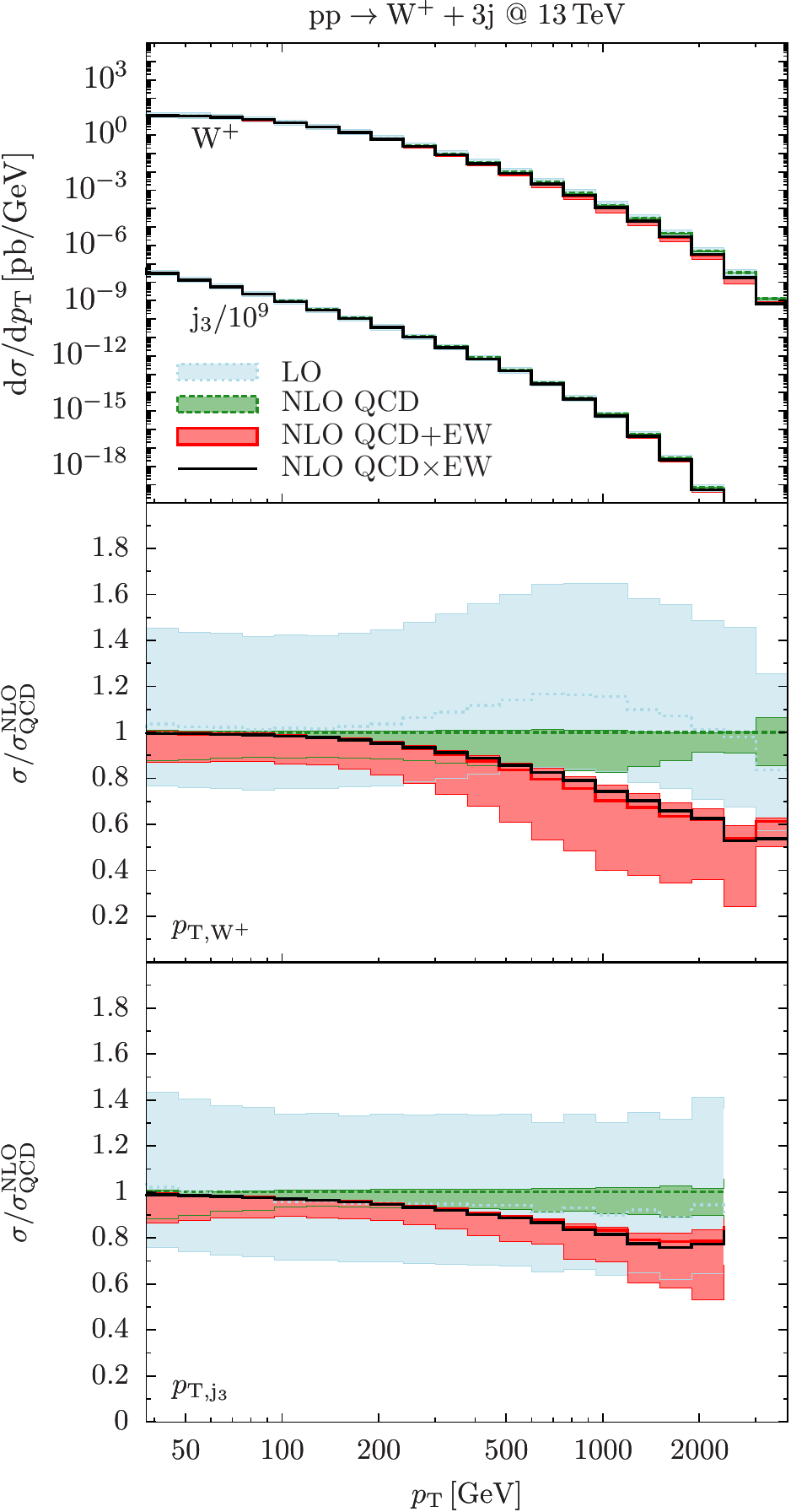}
  \caption{
    \label{Fig:wjj_wjjj}
    NLO QCD and EW corrections to the $W$-boson and subleading and third 
    leading jet transverse momenta, respectively, in $pp\to W^+jj$ (left) 
    and $pp\to W^+jjj$ (right) at 13\,TeV at the LHC. 
  }
\end{figure}

Next-to leading order electroweak corrections to (off-shell) 
vector boson ($V=W, Z, \gamma$) plus multijet production exhibit a rich 
internal structure. Apart from the occurrence of loop diagrams with 
multiple (different) internal masses, interferences from diagrams of 
different order in the QCD and EW couplings, $g_s$ and $e$, are encountered 
as soon as four external quarks are present in the process. Thus, QCD 
and EW corrections cannot solely be classified as corrections due to real 
and virtual parton or photon emissions, respectively, but have to be 
defined through a power counting of the couplings involved: given 
a Born process at $\order(\alphaS^n\alpha^m)$, NLO QCD corrections are 
defined as corrections of $\order(\alphaS^{n+1}\alpha^m)$ while NLO EW 
corrections are of $\order(\alphaS^n\alpha^{m+1})$. Consequently, NLO 
EW real corrections also involve processes with no external photons but 
an emergent pair of quarks or a gluon.

To calculate the NLO QCD and EW corrections to $W+n\,\text{jets}$ 
($n=1,2,3$) we use the fully automated frameworks of \MunichOpenLoops 
and \SherpaOpenLoops. Therein, \OpenLoops \cite{Cascioli:2011va,hepforge} provides 
the renormalised one-loop matrix elements while either \Munich or \Sherpa 
\cite{Gleisberg:2007md,Gleisberg:2008ta} arrange for phase space integration 
and process management. While \Munich uses \OpenLoops matrix elements 
for the Born and real emission matrix elements as well as in its 
Catani-Seymour \cite{Catani:1996vz,Catani:2002hc} subtraction, \Sherpa 
employs its own internal matrix element generators 
\Amegic \cite{Krauss:2001iv} and \Comix \cite{Gleisberg:2008fv}. 
This duality of approaches is used to provide independent cross checks 
of the results. \OpenLoops uses the \Collier \cite{Denner:2016kdg} tensor 
reduction library, employing \CutTools \cite{Ossola:2007ax} and \OneLoop 
\cite{vanHameren:2010cp} for re-evaluating unstable phase space points.

Fig.\ \ref{Fig:wj} presents the NLO QCD and NLO EW corrections on the 
transverse momentum of the $W$ boson and the leading jet in events 
with at least one jet. The inclusive corrections are shown on the left. 
As can be seen, the NLO QCD corrections alone amount to an increase in 
the cross section of several hundred percent at large $\pT$ \cite{Rubin:2010xp}. This 
effect originates in the opening of an alternative channel of production 
of a $W$ boson in association with two jets encountered in the real 
corrections: a hard and nearly back-to-back dijet system radiates a 
comparably soft $W$ boson. As these configurations can be interpreted 
as a NLO real EW boson emission corrections to dijet production, it inherits 
its large initial cross section and dominates in the high-$\pT$ region 
of the leading jet, but not of the $W$ boson. Consequently, its accuracy is reduced to be 
effectively leading order. Similar observations can be made for the NLO 
EW corrections. This artefact is best remedied by a multijet 
merged approach, treating the troublesome dijet plus $W$ configurations 
at next-to leading order as well. In the absence of such a merging in a 
fixed-order calculation the offending contributions can simply be vetoed, 
e.g.\ by requiring the azimuthal separation of the two leading jets to be 
at most $\tfrac{3\pi}{4}$. The results for such a vetoed calculation are 
shown on the right hand side of Fig.\ \ref{Fig:wj}. Here, modest NLO QCD 
and typical Sudakov-type NLO EW corrections are recovered.

NLO QCD and EW corrections for the transverse momentum and the second 
and third leading jet in the higher multiplicity $W+2j$ and $W+3j$ 
events, respectively, are displayed in Fig.\ \ref{Fig:wjj_wjjj}. As 
the Sudakov-type NLO EW corrections in the TeV regime become smaller 
for each subleading jet, so they do for the $W$ boson itself. The 
results of \cite{Kallweit:2014xda} using on-shell bosons were 
quantitatively reproduced with an off-shell calculation using the full 
leptonic final state in \cite{Kallweit:2015dum}.

\begin{figure}[t]
  \centering
  \includegraphics[width=0.47\textwidth]{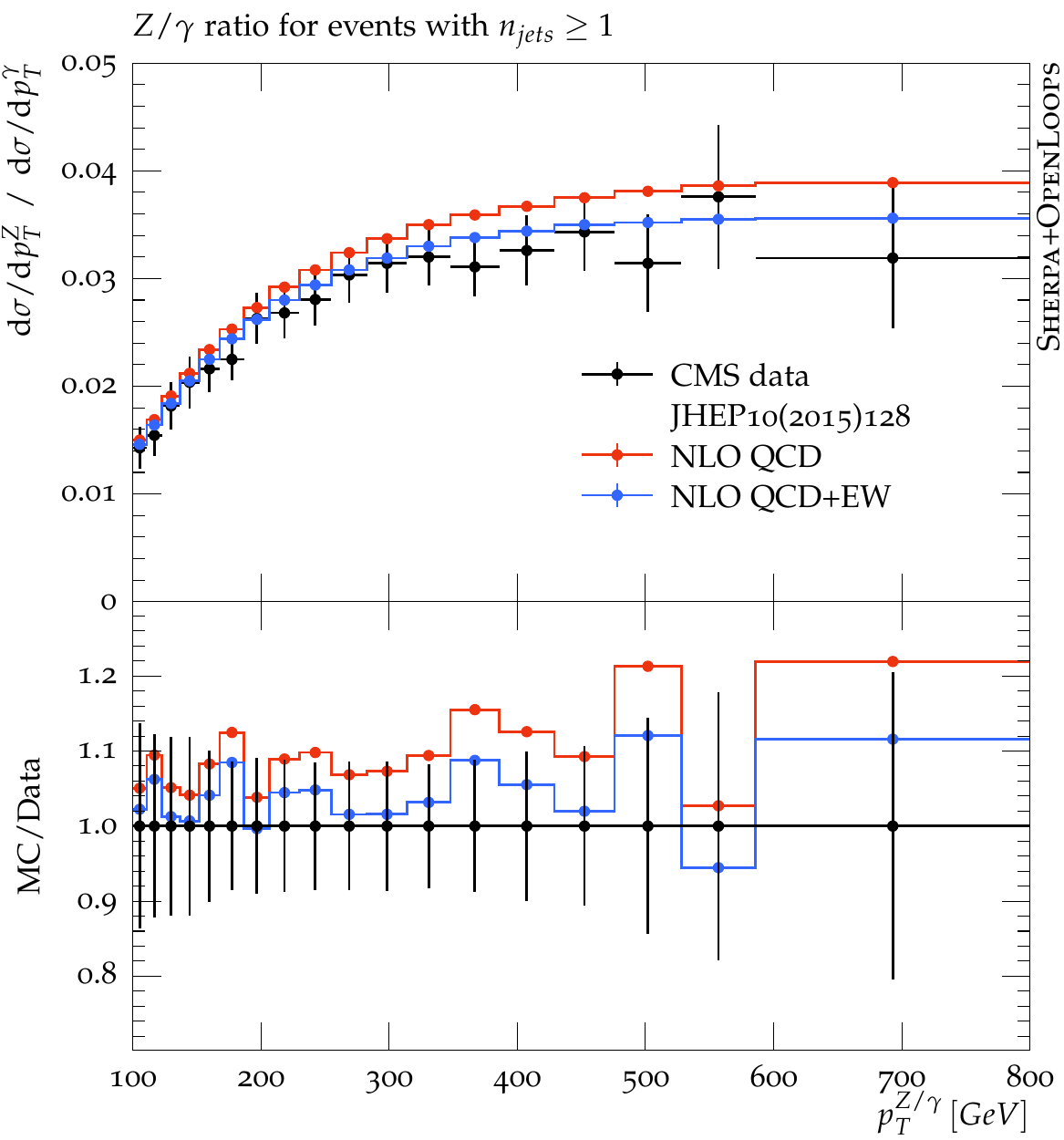}\hfill
  \includegraphics[width=0.47\textwidth]{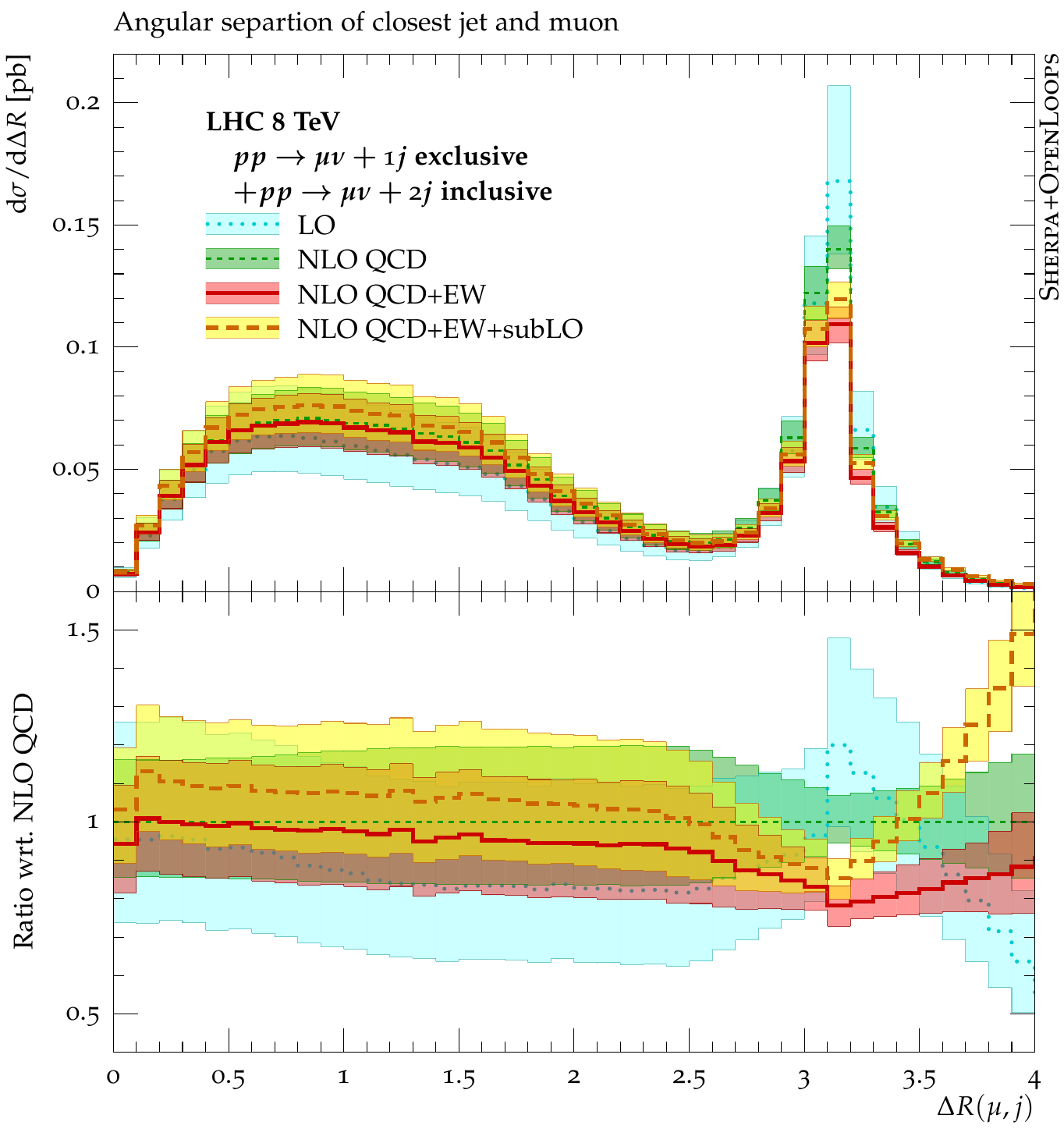}
  \caption{
    \label{Fig:zgamma}
    \textit{Left}: 
    Ratio of the transverse momentum of the reconstructed $Z$-boson and 
    the photon in $pp\to\ell^+\ell^-j$ and $pp\to\gamma j$, respectively, 
    compared to data taken by the CMS collaboration at the LHC at 8\,TeV 
    \cite{Khachatryan:2015ira}.
    \textit{Right}: 
    Angular separation of the muon and the closest jet in $pp\to W+\text{jets}$ 
    production with $\pT^{j_1}>500\,\text{GeV}$ at the LHC at 8\,TeV, 
    a comparison to data can be found in \cite{Wu:2016ichep}.
  }
\end{figure}

These results were then directly used to produce theoretical 
predictions for the $Z/\gamma$-ratio in dependence on the (reconstructed) 
boson transverse momentum \cite{Kallweit:2015fta,Khachatryan:2015ira,
  Badger:2016bpw} in events with at least one jet accompanying the 
reconstructed $Z$ boson or the photon, cf.\ Fig.\ \ref{Fig:zgamma} (left). As 
can be seen, the pure NLO QCD calculation predicts a slightly too large 
ratio. The electroweak correction, acting differently in both processes 
due to the different EW charge of the $Z$ boson as compared to the photon, 
lower the ratio and improve the data description significantly.

Similarly, analyses focussed on finding $W$ boson emissions off high-$\pT$ 
jets \cite{Krauss:2014yaa}, simplified for the limited reach at 8\,TeV, 
primarily measure the angular separation of the muon and the closest jet. 
As in this observables the region $\Delta R \gtrsim \pi$ is already defined 
in $W+1j$ production but $\Delta R \lesssim \pi$ is only populated with $W+2j$ events, 
an exclusive-sums approach (using $\pT^{j_2,\text{cut}}=100\,\text{GeV}$ as 
discriminator) is taken. Fig.\ \ref{Fig:zgamma} (right) displays the NLO QCD+EW prediction for 
this observable including comparatively large subleading Born contributions. 
The yet unpublished study in \cite{Wu:2016ichep} shows excellent agreement 
with data, especially where electroweak corrections 
are sizeable.

\section{Electroweak corrections in particle-level event generation}

The last section demonstrated the size and impact of next-to leading order 
electroweak corrections. However, they need to be incorporated in particle 
level event generators in order to be directly applied in experimental 
analyses. As a proper generic NLO QCD+EW matching to parton showers is still to be 
formulated, the following outlines an approximation that is designed to 
capture the dominant Sudakov-type corrections in the TeV regime as well 
as non-logarithmic ones over the whole phase space.

\begin{figure}[t]
  \includegraphics[width=0.47\textwidth]{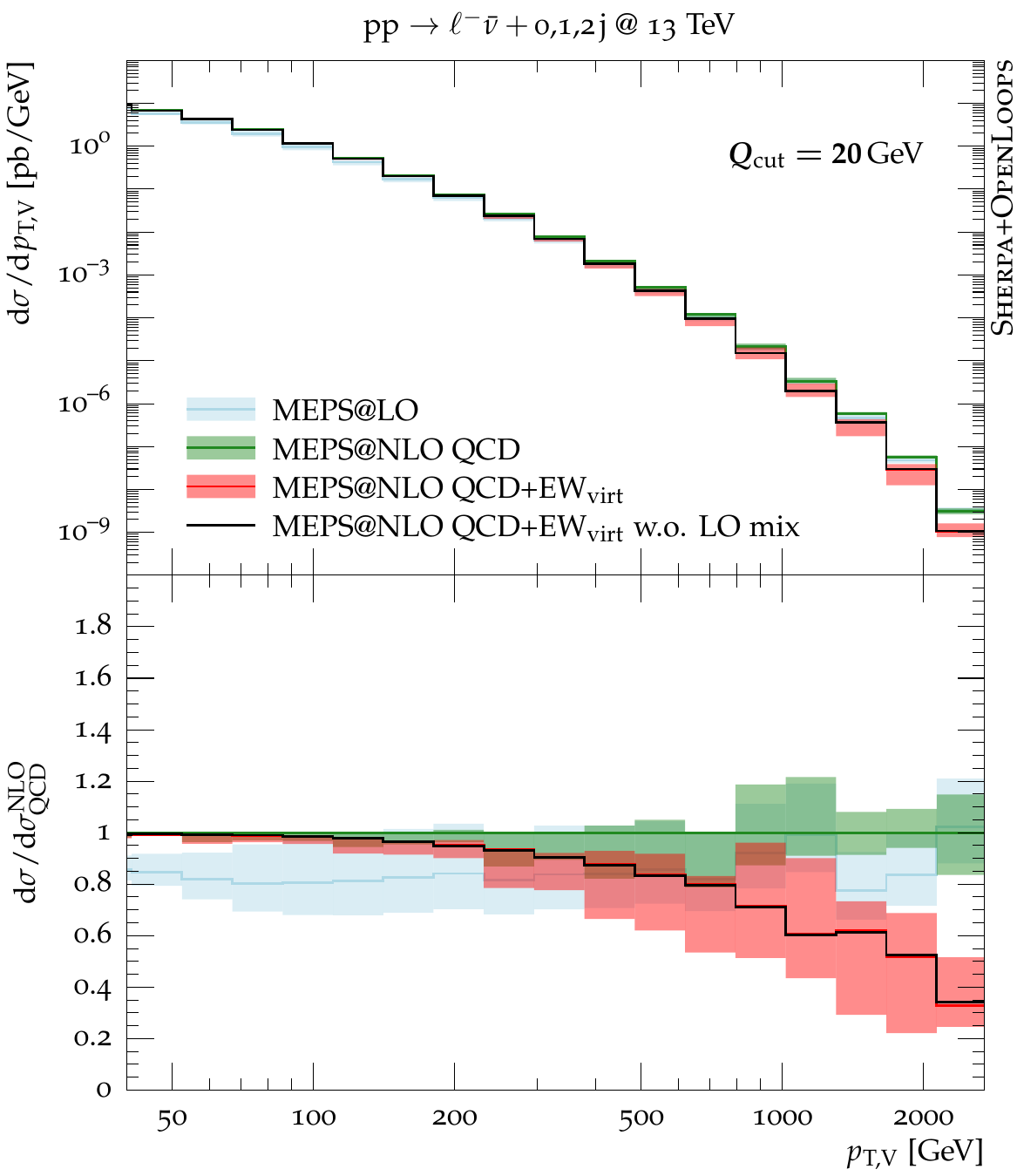}\hfill
  \includegraphics[width=0.47\textwidth]{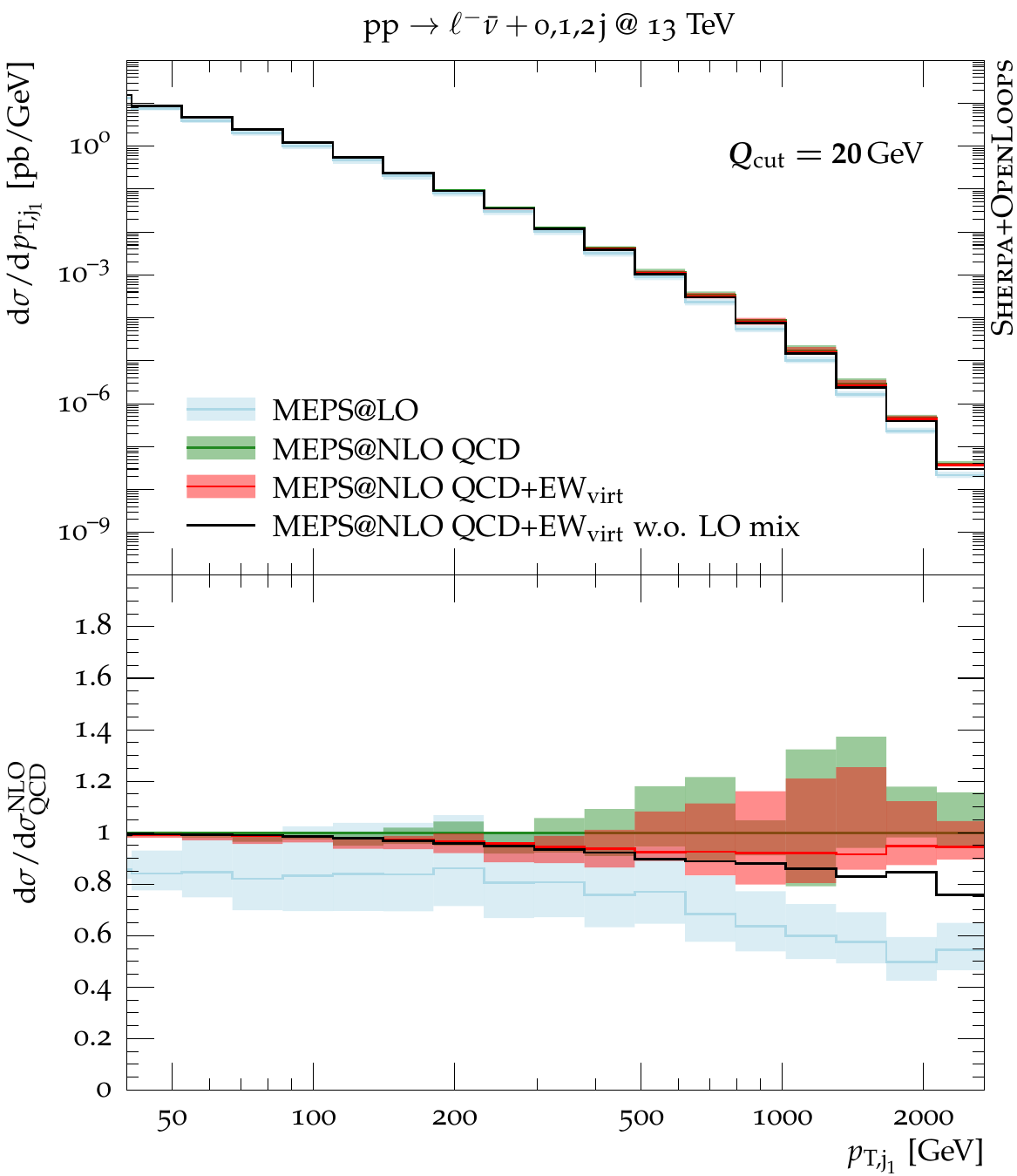}
  \caption{
    \label{Fig:mepsnlo-ewvirt}
    Incorporation of approximate NLO EW corrections into the standard 
    NLO QCD multijet merging for the transverse momenta of the reconstructed 
    $W$-boson (left) and leading jet (right) in $pp\to \ell^-\bar{\nu}_\ell+
    \text{jets}$ at 13\,TeV at the LHC. 
  }
\end{figure}

\begin{figure}[t!]
  \includegraphics[width=0.47\textwidth]{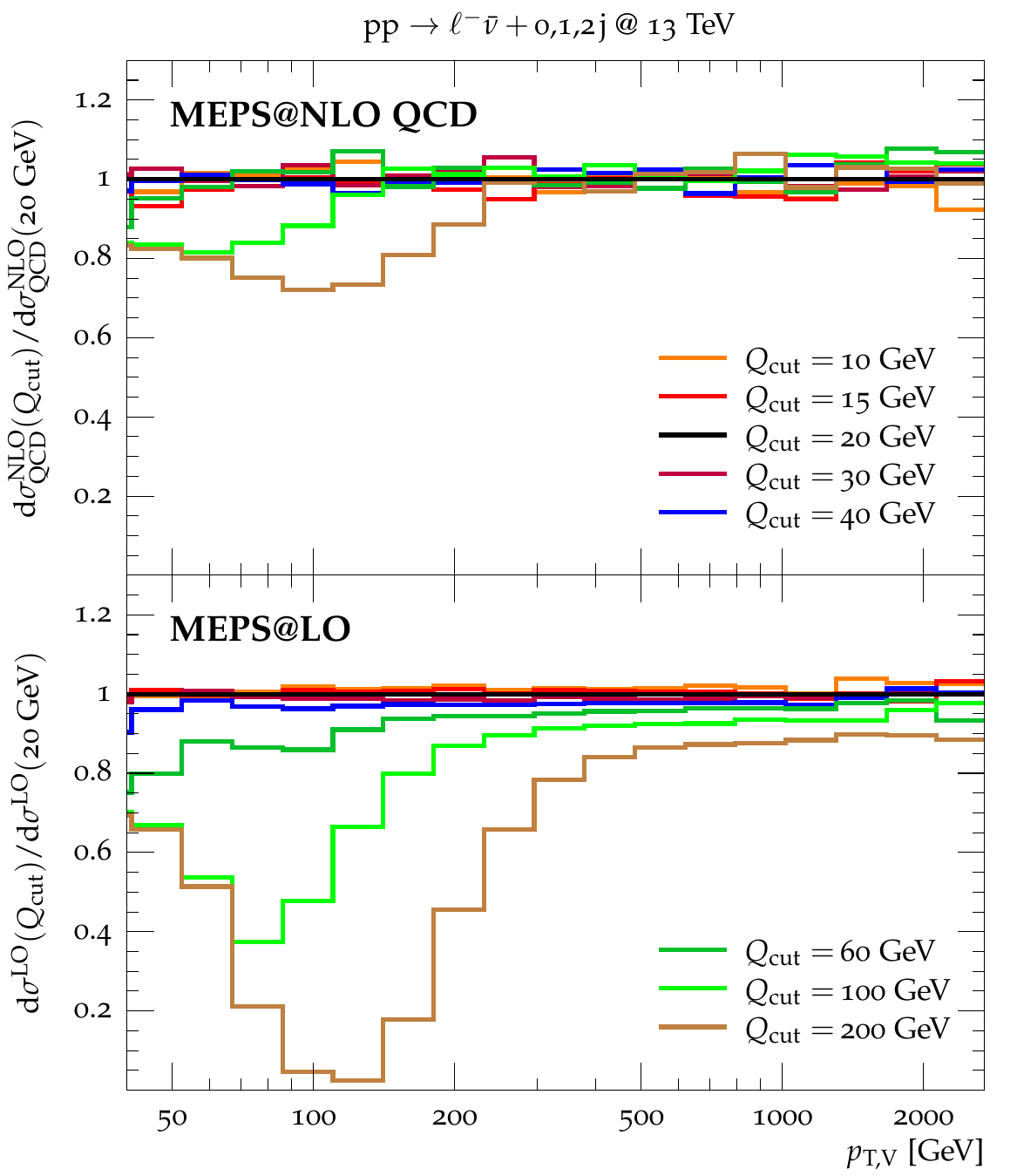}\hfill
  \includegraphics[width=0.47\textwidth]{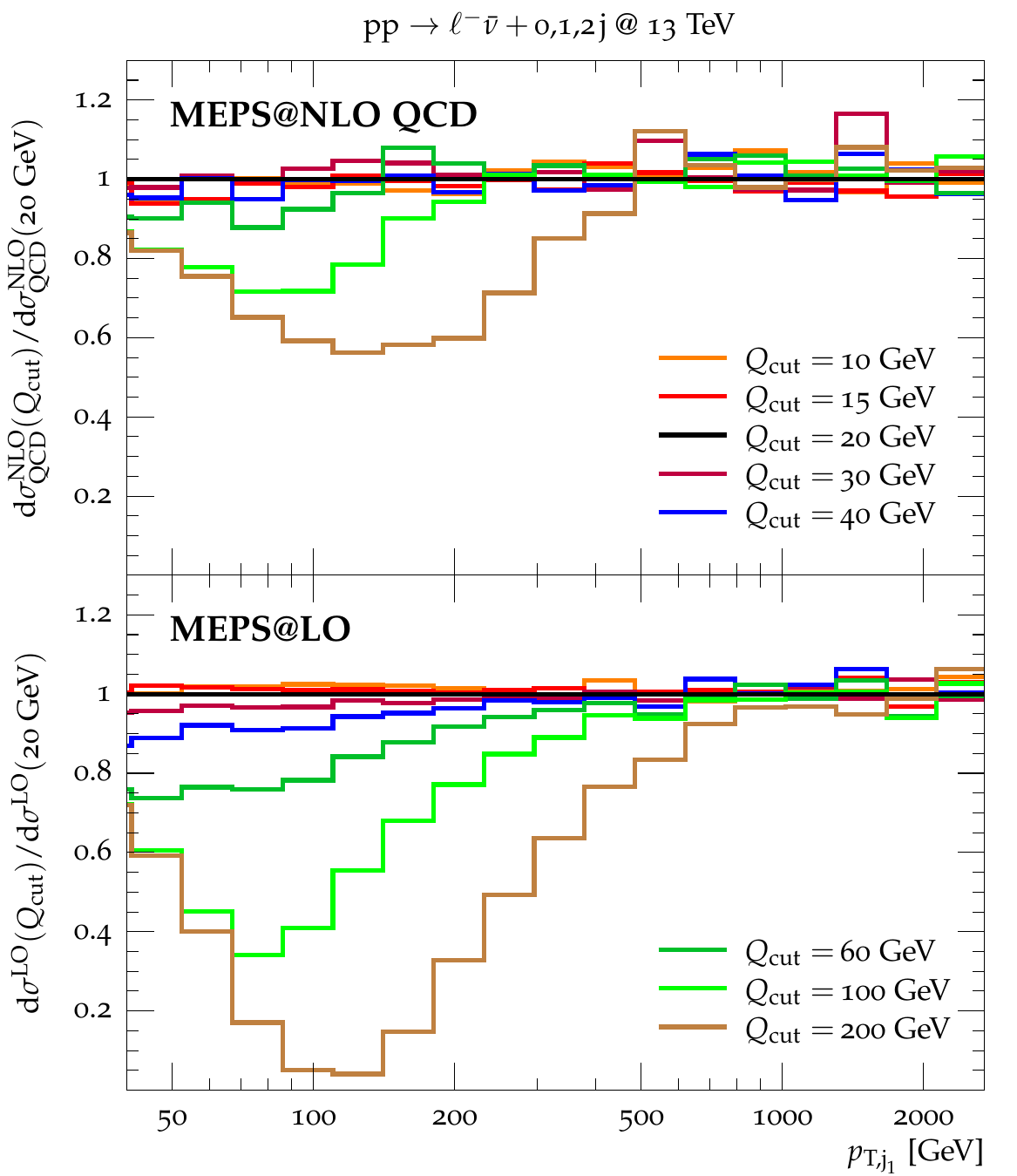}
  \caption{
    \label{Fig:mepsnlo-ewvirt-qcutvar}
    Dependence of the reconstructed $W$-boson (left) and leading jet (right) 
    transverse momenta on the merging parameter $\Qcut$ in the LO (bottom) 
    and NLO (top) QCD multijet merged calculation for $pp\to \ell^-\bar{\nu}_\ell+
    \text{jets}$ at 13\,TeV at the LHC.
  }
\end{figure}

The basis for this approximation is \Sherpa's \MEPSatNLO \cite{Hoeche:2009rj,
  Hoeche:2010kg,Hoeche:2012yf,Gehrmann:2012yg} multijet merging algorithm 
at NLO QCD accuracy. As it bases on an \MCatNLO-type approach \cite{Frixione:2002ik,
  Frixione:2003ei,Hoeche:2011fd,Hoeche:2012ft,Hoeche:2012fm,Hoche:2010pf} 
for matching the NLO QCD computation to the parton shower 
\cite{Schumann:2007mg}, its $\overline{\mr{B}}$-function, labelled 
$\overline{\mr{B}}_\text{QCD}$ in the following, can be extended for the 
$n$ parton final state to
\begin{equation}\label{eq:mepsatnloewa}
  \begin{split}
    \overline{\mr{B}}_{n,\text{QCD+EW}_\text{virt}}(\Phi_{n})=
     \overline{\mr{B}}_{n,\text{QCD}}(\Phi_{n})
        +\mr{V}_{n,\text{EW}}(\Phi_{n})
        +\mr{I}_{n,\text{EW}}(\Phi_{n})
        +\mr{B}_{n,\text{mix}}(\Phi_{n})\;.
  \end{split}
\end{equation}
Herein, $\mr{V}_{n,\text{EW}}(\Phi_{n})$ are the renormalised one-loop 
electroweak corrections while $\mr{I}_{n,\text{EW}}(\Phi_{n})$ are the 
approximated real emission corrections integrated over the one-particle 
emission phase space, such that their sum yields a finite result. Optionally, 
subleading Born contributions, $\mr{B}_{n,\text{mix}}$, can be included 
when numerically relevant. This approach neglects the kinematical impact 
of real photon radiation dominant in e.g.\ leptonic observables. It can 
be recovered, however, by making use of standard QED radiation tools 
\cite{Schonherr:2008av} which have been shown to be superior to a fixed 
order calculation for these observables \cite{Badger:2016bpw}. 

Fig.\ \ref{Fig:mepsnlo-ewvirt} now again shows the transverse momentum 
distribution of the $W$ boson on the left hand side and the leading jet 
on the right hand side obtained when calculating $W+\text{jets}$ production 
with the above outlined method including up to two jets at (approximate) 
next-to leading order accuracy. As can be seen, very similar quantitative 
electroweak corrections are recovered in the TeV regime. Further, due to 
the consistent treatment of one and two jet final states the issue of 
potentially large corrections at large transverse momenta has been 
resolved in the merged calculation and the typical behaviour for the 
leading jet is recovered. For this observable, also the subleading 
Born contributions are of importance as they significantly reduce the impact 
of the electroweak corrections due to their opposite sign.

For the above prediction the standard merging cut of $\Qcut=20\,\text{GeV}$ 
has been used. Following the $\Qcut$-dependence analysis of the method 
provided in \cite{Hoeche:2012yf,Gehrmann:2012yg} large 
uncancelled logarithms could in principle emerge. A direct quantification through 
varying the merging parameter $\Qcut$ yields the results presented in 
Fig.\ \ref{Fig:mepsnlo-ewvirt-qcutvar}, showing negligible dependences for 
a large range of $\Qcut$ values if varied around a sufficiently small central 
value.

\section{Conclusions}

The automation of NLO EW corrections within the \MunichOpenLoops and 
\Sherpa{}+ \OpenLoops frameworks provide indispensable input for many high 
precision or TeV regime analyses at Run II of the LHC. A method for 
incorporation of approximate NLO EW corrections into established multijet 
merging methods at NLO QCD accuracy has been introduced and is publicly 
available within \Sherpa-2.2.1. This formulation allows for particle 
level, and therefore potentially detector simulated, fully realistic 
event simulation. It represents a first, albeit very useful, step 
towards a complete NLO QCD+EW matching and multijet merging.
\vspace*{-1mm}

\paragraph*{Acknowledgements.}
This research was supported in part by the Swiss
National Science Foundation (SNF) under contract PP00P2-128552. 

\vspace*{-3mm}

\bibliographystyle{bib/amsunsrt_mod}
\bibliography{bib/refs}

\end{document}